\documentclass[pra,nobibnotes,aps,showpieces,superscriptaddress,amsfonts,amsmath,float fix,backend=bibtex]{revtex4}
\usepackage[capitalize]{cleveref}
\usepackage{graphicx}
\usepackage{color}
\usepackage{import}
\usepackage{wrapfig}
\usepackage{mathtools}
\usepackage{bm}
\usepackage{amsmath}
\usepackage{amsfonts}
\usepackage{amssymb}

\begin{document}


\title{An alternative approach to several important systems in classical mechanics: energy factorization}

\def\correspondingauthor{\footnote{Corresponding author}}

\author{Karlo Lelas\correspondingauthor{}}
\email{klelas@ttf.unizg.hr}
\affiliation{Faculty of Textile Technology, University of Zagreb, Prilaz baruna Filipovića 28a, 10000 Zagreb, Croatia}

\author{Dario Jukić}
\affiliation{Faculty of Civil Engineering, University of Zagreb,Fra Andrije Kačića-Miošića 26, 10000 Zagreb, Croatia}

\date{\today}

\begin{abstract}
We show how several important classical problems, with positive definite potential energy, can be solved by starting from the factorization of the total mechanical energy using complex numbers. In particular, we derive in a new way exact analytical solutions for: simple harmonic oscillator, vertical projectile motion, motion under a repulsive inverse cube force, and damped harmonic oscillator (with linear damping). We also show how this approach easily yields an excellent approximation of the energy decay and a new approximate analytical solution in the case of a weakly damped harmonic oscillator. Our derivations are suitable for undergraduate physics teaching as an alternative to solving Newton's equations of motion. In addition, we comment on the limitations of our approach, but also on the insights it provides and opportunities for further research.
\end{abstract}

\maketitle

\section{Introduction}
\label{intro}

In typical undergraduate physics textbooks \cite{Cutnell8, UniPhys, Resnick10}, even the solution of a simple harmonic oscillator is not derived by solving Newton's equation of motion, but rather by considering the connection of harmonic oscillations with uniform circular motion. The situation is even worse with a damped harmonic oscillator, for which solutions are typically presented without any derivation \cite{Cutnell8, UniPhys, Resnick10}. The reason for this is that Newton's equation of motion is a second-order differential equation, which is mathematically demanding for first-year undergraduates to solve. Therefore, alternative approaches are needed that lead to exact solutions for these and other interesting systems, using simpler mathematics. The aim of this paper is to offer a new approach to the analysis of the dynamics of several important classical systems, based on physics and mathematics that are well understood by first-year undergraduates.

If we consider one-dimensional dynamics (i.e. motion along the $x$-axis only) of mass $m$ under the influence of conservative forces, the law of conservation of total mechanical energy is given by \cite{UniPhys}
\begin{equation}
\frac{mv^2}{2}+U(x)=E\,,
\label{ZOE}
\end{equation}
where $v=dx/dt$ is the velocity. The first term in \eqref{ZOE} is the kinetic energy, and $U(x)$ is the potential energy corresponding to the conservative force $F(x)=-dU(x)/dx$. For conservative forces, equation \eqref{ZOE} can be derived without knowing the explicit form of functions $x(t)$ and $v(t)$, i.e. on general grounds, using the work-energy theorem \cite{UniPhys}. Here we analyze the dynamics in several important cases with positive-definite potential energy, i.e. with $U(x)\geq0$, and the starting point of our analysis is equation \eqref{ZOE}. In that case, we can define an auxiliary real-valued function $u(x)$ such that
\begin{equation}
U(x)=u^2(x)
\label{pomocna}
\end{equation}
holds. We note here that the auxiliary function $u(x)$ can generally have positive and negative values and be equal to zero. Thus, $|u(x)|=\sqrt{U(x)}$ but $u(x)\neq\sqrt{U(x)}$ in general, since the square root function is defined to return only non-negative real numbers. Using \eqref{pomocna}, we can factor \eqref{ZOE} as
\begin{equation}
\left(\sqrt{\frac{m}{2}}v+i\,u(x)\right)\left(\sqrt{\frac{m}{2}}v-i\,u(x)\right)=E\,,
\label{ZOE1}
\end{equation}
where $i=\sqrt{-1}$, and $E>0$ holds (here we exclude the case $E=0$ from consideration, since, due to constraint $U(x)\geq0$, it can be fulfilled only in the trivial case with $v(t)=0$ $\forall t$ and $U(x)=0$ $\forall x$, such as a free particle at rest). The expressions in parentheses on the left-hand side of \eqref{ZOE1} are complex conjugates. Therefore, the expression in the first parentheses can be written as
\begin{equation}
\sqrt{\frac{m}{2}}v(t)+i\,u(x(t))=\sqrt{E}\,e^{i\phi(t)}\,,
\label{prva}
\end{equation}
and the expression in the second parentheses as
\begin{equation}
\sqrt{\frac{m}{2}}v(t)-i\,u(x(t))=\sqrt{E}\,e^{-i\phi(t)}\,,
\label{druga}
\end{equation}
where we emphasized that position and velocity depend on time, and introduced an unknown function $\phi(t)$, i.e. the time-dependent phase. Since $e^{\pm i\phi(t)}=\cos\phi(t)\pm i\sin\phi(t)$, by adding equations \eqref{prva} and \eqref{druga} we get
\begin{equation}
v(t)=\sqrt{\frac{2E}{m}}\cos\phi(t)\,,
\label{brzina}
\end{equation}
and by subtracting them we get
\begin{equation}
u(x(t))=\sqrt{E}\sin\phi(t)\,.
\label{polozaj}
\end{equation}
Equations \eqref{brzina} and \eqref{polozaj} are the cornerstones of our approach. In what follows, we use them to determine $x(t)$ for several fundamental systems in classical physics.

The physical meaning of phase $\phi(t)$ is that it tells us how the total energy $E$ is distributed between the potential energy and the kinetic energy at any given instant, i.e. $U(x(t))=E\sin^2\phi(t)$ and $mv^2(t)/2=E\cos^2\phi(t)$. Furthermore, we can easily see that its tangent is related to the square root of the potential to kinetic energy ratio, i.e.
\begin{equation}
\tan\phi(t)=\frac{u(x(t))}{\sqrt{mv^2(t)/2}}=\text{sgn}[u(x(t))]\sqrt{\frac{U(x(t))}{mv^2(t)/2}}\,
\label{tan}
\end{equation}
is valid, where we use the sign function $\text{sgn}[u(x(t))]$ (equal to $1$ for $u(x(t))>0$, $-1$ for $u(x(t))<0$, and $0$ for $u(x(t))=0$) after the second equality to take into account that $u(x(t))$ and $\sqrt{U(x(t))}$ can differ up to a sign in general. 

We note here that similar approaches have already been introduced in the context of harmonic oscillators \cite{Gauthier2004, Tisdell2019, AJP2025HO}, but none of them start from factorization of total mechanical energy using basic knowledge of complex numbers. For example, in a recent paper \cite{AJP2025HO}, the starting point is a system of coupled first-order differential equations for position and momentum. The equations are then decoupled by introducing a new function which is a linear combination of position and momentum with complex coefficients. Thus, the process of decoupling the equations led to the introduction of complex numbers. After solving the uncoupled equations for the new function and its complex conjugate, linear combinations of the new function and its complex conjugate were used to obtain solutions for position and momentum as functions of time, and it was subsequently shown that the energy of the simple harmonic oscillator is conserved \cite{AJP2025HO}. 

In an even more recent paper \cite{Zengel2026}, along with a simple harmonic oscillator, systems with a wider class of positive definite potential energies were considered but starting from more demanding theoretical concepts, such as: action, Lagrangian, Euler-Lagrange equations, canonical transformations and the Hamilton–Jacobi theory. An alternative Lagrangian has been formulated in \cite{Zengel2026}, whose Euler-Lagrange equation is a first-order differential equation for the variable $\left(\sqrt{m/2}\,v(t)\pm i\sqrt{U(x(t))}\right)$, which further leads to equations similar to our equations \eqref{brzina} and \eqref{polozaj}.

Our starting point is easier to understand. In particular, in the case of a simple harmonic oscillator (or any other conservative system), it is easy to show that energy is conserved, without knowing how position and velocity (momentum) depend on time, and therefore our starting point is the equation \eqref{ZOE}. The steps from \eqref{ZOE} to the equations \eqref{brzina} and \eqref{polozaj} are straightforward and easy to understand, since anyone with elementary knowledge of complex numbers knows that $a^2+b^2$, where $a$ and $b$ are real numbers, can be written as $(a+i\,b)(a-i\,b)$ and that complex conjugates have the same modulus but differ in the sign of the phase. Furthermore, we show in what follows that a damped harmonic oscillator, i.e. a dissipative system, can be solved exactly with our approach. In contrast, dissipative systems are not discussed or mentioned in \cite{Zengel2026}, perhaps due to the complexity of the formalism used.

We note here that one can come to equations \eqref{brzina} and \eqref{polozaj} even without using complex numbers, which can be useful, e.g. if students are not yet familiar with Euler's representation of a complex number. Using \eqref{pomocna}, the law of conservation of total mechanical energy can be rewritten as
\begin{equation}
\frac{m}{2E}v^2(t)+\frac{1}{E}u^2(x(t))=1\,.
\label{ZOE11}
\end{equation}
In \eqref{ZOE11}, we have two positive definite time-dependent terms that give $1$ when added together. Now we just need to remember that the identity $\sin^2\phi(t)+\cos^2\phi(t)=1$ holds for any function $\phi(t)$. Thus, we can simply take that the first term on the left hand side of \eqref{ZOE11} is equal to $\cos^2\phi(t)$ and the second equal to $\sin^2\phi(t)$. 



\section{Simple harmonic oscillator}
\label{SHO}

In the case of potential energy $U(x)=k\,x^2/2$, i.e. the case with $u(x)=(\sqrt{k/2})x$ and force $F(x)=-k x$, equation \eqref{polozaj} becomes
\begin{equation}
x(t)=\sqrt{\frac{2E}{k}}\sin\phi(t)\,.
\label{xHO}
\end{equation}
The first derivative of \eqref{xHO} with respect to time is
\begin{equation}
\frac{dx(t)}{dt}=\sqrt{\frac{2E}{k}}\left(\frac{d\phi(t)}{dt}\right)\cos\phi(t)\,.
\label{xHOdt}
\end{equation}
The unknown function $\phi(t)$ must be such that \eqref{xHOdt} corresponds to the velocity \eqref{brzina}, i.e. if we equate the right hand sides of \eqref{xHOdt} and \eqref{brzina} we get 
\begin{equation}
\frac{d\phi(t)}{dt}=\sqrt{\frac{k}{m}}\,.
\label{dPhi}
\end{equation}
Thus, we get a simple first order differential equation which can be solved using the separation of variables method, i.e. by solving the integral $$\int_{\phi(0)}^{\phi(t)}d\phi(t')=\omega_0\int_0^tdt'\,,$$ where we introduce the standard notation $\omega_0=\sqrt{k/m}$. Of course, even without knowing integral calculus, it is easy to argue using only elementary knowledge about the first derivatives that the solution to equation \eqref{dPhi} is
\begin{equation}
\phi(t)=\omega_0t+\phi_0\,,
\label{fazaHO}
\end{equation}
where $\phi_0=\phi(0)$ is the initial phase. Thus, in the case of simple harmonic oscillator the solution is
\begin{equation}
x(t)=\sqrt{\frac{2E}{k}}\sin\left(\omega_0t+\phi_0\right)\,.
\label{solutionHO}
\end{equation}
The amplitude $\sqrt{2E/k}$ and the initial phase are easily obtained from initial conditions $x(0)=x_0$ and $v(0)=v_0$. For example, if the dynamics starts with the initial conditions $x_0>0$ and $v_0=0$, we have $E=kx_0^2/2$ and $\phi_0=\pi/2$, and \eqref{solutionHO} becomes $x(t)=x_0\cos(\omega_0t)$. 

We note here that this is an example of a system for which the corresponding auxiliary function $u(x)$ can have a different sign than $\sqrt{U(x)}$, since here $u(x)\in\left[-\sqrt{E},\sqrt{E}\right]$. It was proposed in \cite{Zengel2026} to think about variable $R(t)=\sqrt{m/2}\,v(t)+i\sqrt{U(x(t))}$ as a phasor of constant length $\sqrt{E}$ rotating in a complex plane defined by the real axis $\sqrt{m/2}\,v$ and the imaginary axes $\sqrt{U(x)}$. It was stated that the phasor $R(t)$ rotates at a constant rate $\omega_0$ in the case of a harmonic oscillator \cite{Zengel2026}. Strictly speaking, in the case of a harmonic oscillator, the phasor defined in this way would not rotate at a constant rate $\omega_0$, but would rather change the direction of its rotation (instantly) each time the system passes through the equilibrium position, since by construction $R(t)$ has to stay in the upper half-plane (with positive imaginary values).

Therefore, for a phasor-like description of the energy, it is practical to introduce an auxiliary function $u(x)$, for which $u^2(x)=U(x)$ holds, as we have done here. The phasor $\tilde{R}(t)=\sqrt{m/2}\,v(t)+iu(x(t))$ has a constant magnitude $\sqrt{E}$ and, in the case of a harmonic oscillator, rotates counter-clockwise with an angular velocity $d\phi(t)/dt=\omega_0$ in the complex $\left(\sqrt{m/2}\,v, u(x)\right)$ plane.

\section{Homogeneous gravitational potential energy}
\label{gravity}

In the case of potential energy $U(x)=mgx$, where we take $x=0$ as the reference level (ground level), and where the positive direction of the $x$-axis is upward (i.e. $x\geq0$ during the vertical projectile motion), $u(x)=\sqrt{U(x)}$ holds and we can rewrite equation \eqref{polozaj} as   
\begin{equation}
x(t)=\frac{E}{mg}\sin^2\phi(t)\,.
\label{xG}
\end{equation}
We can solve this case even more simply than the harmonic oscillator. Namely, since the associated conservative force is of constant magnitude, i.e. $F(x)=-mg$, we know that this is a motion with constant acceleration $a=-g$. Therefore, the instantaneous and average accelerations are equal, i.e. $dv(t)/dt=\Delta v/\Delta t=-g$, where $\Delta v=v(t)-v_0$, $\Delta t=t$, and $v_0$ is the initial velocity (along the $x$-axis). Thus, one can easily conclude, using only basic definitions of instantaneous and average accelerations \cite{UniPhys}, that the instantaneous velocity is given by
\begin{equation}
v(t)=v_0-gt\,.
\label{velocity}
\end{equation}
If we equate the right hand sides of \eqref{brzina} and \eqref{velocity} we get
\begin{equation}
\cos\phi(t)=\sqrt{\frac{m}{2E}}\left(v_0-gt\right)\,.
\label{cos}
\end{equation}
If we use \eqref{cos} and identity $\sin^2\phi(t)=1-\cos^2\phi(t)$ in \eqref{xG} we easily get
\begin{equation}
x(t)=\frac{E}{mg}-\frac{1}{2g}\left(v_0-gt\right)^2\,.
\label{xG1}
\end{equation}
If we take, e.g., initial conditions $x_0=h$ and $v_0>0$, i.e. $E=mgh+mv_0^2/2$, \eqref{xG1} becomes
\begin{equation}
x(t)=h+v_0t-\frac{g}{2}t^2\,,
\label{xG2}
\end{equation}
which is the more familiar form of the solution \cite{UniPhys}. Using \eqref{cos} we easily get
\begin{equation}
\phi(t)=\arccos\left(\sqrt{\frac{m}{2E}}\left(v_0-gt\right)\right)\,.
\label{fazaG}
\end{equation}
Although in this particular case we did not need the phase $\phi(t)$ to determine $x(t)$, here we give an expression \eqref{fazaG} because it is interesting to consider the behavior of the corresponding phasor and its angular velocity. Thus, a phasor-like description of energy in the complex $(\sqrt{m/2}\,v,u(x))$ plane can be easily established, similarly as in \cite{Zengel2026}. For example, for initial conditions $h=0$ and $v_0>0$, i.e. $E=mv_0^2/2$, we can easily calculate that the projectile reaches the maximal hight at $t=v_0/g$ and falls back to the ground at $t=2v_0/g$. In that case, the relation \eqref{fazaG} gives $\phi(0)=0$, $\phi(v_0/g)=\pi/2$, and $\phi(2v_0/g)=\pi$. Thus, during the upward and downward motion of the projectile, the phasor $\tilde{R}(t)=\sqrt{m/2}\,v(t)+i\sqrt{mgx(t)}$ rotates counter-clockwise from $\phi(0)=0$ to $\phi(2v_0/g)=\pi$. The angular velocity of this rotation is given by the time derivative of \eqref{fazaG}, i.e. 
\begin{equation}
\frac{d\phi(t)}{dt}=\frac{g}{\sqrt{gt(2v_0-gt)}}\,.
\label{kutnaG}
\end{equation}
Thus, $d\phi(t)/dt>0$ for $0\leq t \leq 2v_0/g$, i.e., as we already said, the rotation of the phasor is counter-clockwise. Furthermore, we can see that angular velocity has vertical asymptotes, i.e. tends to infinity at $t=0$ and $t=2v_0/g$, while it has minimal value $d\phi(t)/dt=g/v_0$ at $t=v_0/g$.


\section{Repulsive potential energy with the inverse-cube force law}
\label{repulsive}

Another interesting case is the potential energy $U(x)=K x^{-2}/2$, with $K>0$, i.e. the case with $u(x)=(\sqrt{K/2})x^{-1}$ and the force $F(x)= K x^{-3}$ (notice that the potential barrier is impenetrable at $x=0$, so the motion is restricted to a semi-infinite line, e.g. $x>0$ or $x<0$, depending on initial conditions). A physically relevant situation where such an inverse-cube force arises is when the motion is effectively one-dimensional, for example when a point charge moves along the axis of a fixed electric dipole with fixed orientation. In this case, the potential reduces to a $1/r^2$ form along the dipole axis (see, e.g. Ref. \cite{Griffiths2017}, for the multipole expansion). In this case, equation \eqref{polozaj} becomes
\begin{equation}
\frac{1}{x(t)}=\sqrt{\frac{2E}{K}}\sin\phi(t)\,.
\label{xRep}
\end{equation}
The first derivative of \eqref{xRep} with respect to time is
\begin{equation}
-\frac{1}{x^2(t)}\frac{dx(t)}{dt}=\sqrt{\frac{2E}{K}}\left(\frac{d\phi(t)}{dt}\right)\cos\phi(t)\,.
\label{xRepdt}
\end{equation}
Using \eqref{xRep} in \eqref{xRepdt} we get
\begin{equation}
\frac{dx(t)}{dt}=-\sqrt{\frac{K}{2E}}\left(\frac{1}{\sin^2\phi(t)}\cdot\frac{d\phi(t)}{dt}\right)\cos\phi(t)\,.
\label{xRepdt1}
\end{equation}
The unknown function $\phi(t)$ must be such that \eqref{xRepdt1} corresponds to the velocity \eqref{brzina}. We equate the right hand sides of \eqref{xRepdt1} and \eqref{brzina} and get the differential equation
\begin{equation}
\frac{1}{\sin^2\phi(t)}\cdot\frac{d\phi(t)}{dt}=-\frac{2E}{\sqrt{mK}}\,.
\label{faza0}
\end{equation}
Equation \eqref{faza0} can be solved using separation of variables and integration, i.e. we have to solve
\begin{equation}
\int_{\phi_0}^{\phi(t)}\frac{d\phi(t')}{\sin^2\phi(t')}=-\frac{2E}{\sqrt{mK}}\,t\,.
\label{faza1}
\end{equation}
Integral in \eqref{faza1} is elementary, we get
\begin{equation}
\phi(t)=\arctan\left[\left(\frac{2E}{\sqrt{mK}}\,t+\cot\phi_0\right)^{-1}\right]\,.
\label{fazaRep}
\end{equation}
As an example, we consider again initial conditions $x_0>0$ and $v_0=0$. In this case, $E=K/(2x_0^2)$ and $\phi_0=\pi/2$. Thus, $\cot\phi_0=0$ in \eqref{fazaRep}. Relation \eqref{xRep} becomes
\begin{equation}
x(t)=x_0\left[\sin\left(\arctan\left(x_0^2\sqrt{\frac{m}{K}}\,t^{-1}\right)\right)\right]^{-1}\,.
\label{xRep1}
\end{equation}
We can use the identity $\sin[\arctan(\alpha)]=\alpha/\sqrt{1+\alpha^2}$ in \eqref{xRep1}, and obtain
\begin{equation}
x(t)=\sqrt{\left(\frac{K}{mx_0^2}\right)\,t^2+x_0^2}\,.
\label{xRep2}
\end{equation}

The corresponding phasor $\tilde{R}(t)=\sqrt{m/2}\,v(t)+i\sqrt{K/2}\,x^{-1}(t)$ has a constant magnitude $\sqrt{K/2}\,x_0$ and initial phase $\phi(0)=\pi/2$. The angular velocity is given by the time derivative of \eqref{fazaRep}, i.e. by
\begin{equation}
\frac{d\phi(t)}{dt}=-\frac{x_0^2\sqrt{m/K}}{t^2+x_0^4m/K}\,.
\label{kutnaRep}
\end{equation}
Since $d\phi(t)/dt<0$ for all $t\geq0$, we can conclude that the corresponding phasor rotates clockwise (from the initial phase $\pi/2$ towards the phase $0$), but the magnitude of the angular velocity monotonically decreases with time and the phase asymptotically approaches zero, i.e. $\phi(t\rightarrow\infty)\rightarrow0$.

\section{Central force field: effective potential energy}
\label{central}

It is well known that for a particle of mass $m$ moving in a three-dimensional central force field, i.e.\ under a force $\mathbf{F}(\mathbf{r}) = F(r)\,\hat{\mathbf{r}}$, the dynamics in spherical coordinates reduces to an effective 1D problem for the radial coordinate $r(t)$. Angular momentum $L$ is conserved, and the conservation of energy reads \cite{Goldstein2002}
%
%
%
\begin{equation}
\frac{m}{2}\left(\frac{dr(t)}{dt}\right)^2+U_{\mathrm{eff}}(r)=E\,,
\label{eq:radialeq}
\end{equation}
where the effective potential energy is
\begin{equation}
U_{\mathrm{eff}}(r) = U(r) + \frac{L^{2}}{2m r^{2}}.
\label{eq:Ueff}
\end{equation}
Here, $U(r)$ is the (physical) potential energy in 3D, and the term $L^{2}/(2mr^{2})$ represents the familiar centrifugal barrier. We note here that in the corresponding two-body problem (when the masses $m_1$ and $m_2$ interact through $U(r)$, where $r=|\mathbf{r}_1-\mathbf{r}_2|$), the mass $m$ in \eqref{eq:radialeq} and \eqref{eq:Ueff} is replaced by the reduced mass $\mu = m_{1}m_{2}/(m_{1}+m_{2})$.

In special case when $L=0$, the motion is purely radial and all examples from previous sections (harmonic oscillator, homogeneous gravitational field, inverse-cube force) can be solved in this 3D context. Furthermore, if the potential energy is $U(r)=K/r^{2}$, then
\begin{equation}
U_{\mathrm{eff}}(r)=\frac{K}{r^{2}}+\frac{L^{2}}{2mr^{2}}
= \frac{\alpha}{r^{2}},
\label{eq:Ueff_inverse_square}
\end{equation}
with $\alpha \equiv  K + L^{2}/(2m)$, i.e. the effective potential energy is inverse-squared for any angular momentum $L$, and the analysis of Section \ref{repulsive} applies. 
This is also valid for free particle, i.e. in the case $U(r)=0$ with $L>0$, for which the effective potential energy contains only the centrifugal term $L^2/(2mr^2)$. Of course, direct integration of the equation of motion in the standard approach is also applicable for these cases \cite{Goldstein2002}, therefore our method does not provide a fundamentally new insight in these cases, but rather an alternative derivation.

In addition, we note that in the case $U(r)=k/r$, with $L>0$, the effective potential energy
\[
U_{\mathrm{eff}}(r)=\frac{k}{r}+\frac{L^{2}}{2 m r^{2}}
\]
can be rewritten as 
\begin{equation}
U_{\mathrm{eff}}(r)=\left(\frac{L}{\sqrt{2m}\,r}+kL\sqrt{\frac{m}{2}}\right)^2-\frac{mk^2L^2}{2}\,.
\label{kvadratura}
\end{equation}
Using \eqref{kvadratura} in \eqref{eq:radialeq}, we get
\begin{equation}
\frac{m}{2}\left(\frac{dr(t)}{dt}\right)^2+\left(\frac{L}{\sqrt{2m}\,r}+kL\sqrt{\frac{m}{2}}\right)^2=\tilde{E}\,,
\label{kvadratura1}
\end{equation}
where $\tilde{E}\equiv E+mk^2L^2/2$. If $k>0$ (e.g. for Coulomb repulsion), $\tilde{E}$ is a positive definite quantity, since $E>0$ necessarily holds. If $k<0$ (e.g. gravitational potential energy or Coulomb attraction), $E$ can be positive or negative, but even for $E<0$ we can still have $\tilde{E}>0$ depending on the value of $L^2$. Thus, there are interesting cases with $U(r)=k/r$ for which our approach can be used in principle, i.e. we can factor the left hand side of the equation \eqref{kvadratura1} as we did with equation \eqref{ZOE}. Unfortunately, although in these cases the associated phase integral can be solved analytically, the resulting expressions are such that $\phi(t)$, and consequently $r(t)$, cannot be obtained in a closed form. The similar issue is present in the standard approach to analysis of dynamics in the case of potential energy $U(r)=k/r$, i.e. in the standard treatment of the Kepler problem, $r(t)$ cannot be expressed in closed form \cite{Goldstein2002}.

\section{Dissipative system: damped harmonic oscillator}
\label{DHO}

\subsection{Exact solution and exact energy decay}
\label{DHO1}

In the case of mass $m$ oscillating under the influence of conservative restoring force $F(x)=-kx$ and nonconservative (dissipative) damping force $F_d(v)=-b\,v$, where $v=dx/dt$, the exact analytical solutions to the equation of motion are well known and studied in detail, e.g. see \cite{UniPhys, Morin1, Dourmashkin, Berkeley}, but the full derivation of the solutions is often omitted even in textbooks at a more advanced undergraduate level, e.g., see \cite{Berkeley}. Here we show that our approach can be applied in this case as well, as an alternative to solving the equation of motion. 

The total mechanical energy decays over time, due to damping, i.e. $E(t)=mv^2(t)/2+kx^2(t)/2$ is a function of time, but we can still proceed with the same factorization of energy as in examples with conservative forces only. Thus, equations \eqref{brzina} and \eqref{polozaj} now become
\begin{equation}
v(t)=\sqrt{\frac{2E(t)}{m}}\cos\phi(t)\,
\label{brzinaDHO}
\end{equation}
and 
\begin{equation}
x(t)=\sqrt{\frac{2E(t)}{k}}\sin\phi(t)\,.
\label{polozajDHO}
\end{equation}
In addition to equations \eqref{brzinaDHO} and \eqref{polozajDHO}, the equation of energy dissipation rate $dE(t)/dt=-bv^2(t)$ must be satisfied in this case as well, e.g. see \cite{UniPhys}. Using \eqref{brzinaDHO}, the energy dissipation rate can be written in terms of $E(t)$ and $\phi(t)$, i.e.
\begin{equation}
\frac{dE(t)}{dt}=-\frac{2b}{m}E(t)\cos^2\phi(t)\,
\label{dEDHO}
\end{equation}
must be valid.

The first derivative of \eqref{polozajDHO} with respect to time is
\begin{equation}
\frac{dx(t)}{dt}=\sqrt{\frac{1}{2kE(t)}}\sin(\phi(t))\frac{dE(t)}{dt}+\sqrt{\frac{2E(t)}{k}}\cos(\phi(t))\frac{d\phi(t)}{dt}\,.
\label{xDHOdt}
\end{equation}
We insert \eqref{dEDHO} into \eqref{xDHOdt} and get
\begin{equation}
\frac{dx(t)}{dt}=\sqrt{\frac{2E(t)}{m}}\cos\phi(t)\left(-\frac{b}{2m\omega_0}\sin(2\phi(t))+\frac{1}{\omega_0}\frac{d\phi(t)}{dt}\right)\,,
\label{xDHOdt1}
\end{equation}
where we used $\omega_0=\sqrt{k/m}$. The unknown function $\phi(t)$ must be such that \eqref{xDHOdt1} corresponds to the velocity \eqref{brzinaDHO}, i.e. if we equate the right hand sides of \eqref{xDHOdt1} and \eqref{brzinaDHO} we get
\begin{equation}
-\frac{b}{2m\omega_0}\sin(2\phi(t))+\frac{1}{\omega_0}\frac{d\phi(t)}{dt}=1\,.
\label{fazaDHO}
\end{equation}
Thus, the phase is obtained by solving the integral
\begin{equation}
\int_{\phi_0}^{\phi(t)}\frac{d\phi(t')}{1+\frac{\gamma}{\omega_0}\sin(2\phi(t'))}=\omega_0\,t\,,
\label{fazaDHOintegral}
\end{equation}
where we introduced $\gamma=b/(2m)$, i.e. a commonly used notation for the damping coefficient. The value of the ratio $\gamma/\omega_0$ defines three damping regimes \cite{Berkeley}, i.e. the underdamped regime for $0<\gamma/\omega_0<1$, the critically damped regime for $\gamma/\omega_0=1$, and the overdamped regime for $\gamma/\omega_0>1$. 

In what follows, we determine the phase $\phi(t)$ and the solution $x(t)$ for the underdamped case with initial conditions $x_0>0$ and $v_0=0$. For this choice of initial conditions $\phi_0=\pi/2$. We solve the integral \eqref{fazaDHOintegral} in appendix \ref{appendixA}, and get (see \eqref{fazaAppendix})   
\begin{equation}
\phi(t)=\arctan\!\left[-\frac{\omega}{\omega_0}\,\cot(\omega t)-\frac{\gamma}{\omega_0}\right]\,
\label{PhiDHO}
\end{equation}
where $\omega=\sqrt{\omega_0^2-\gamma^2}$. 
%
%
We are now in a position to obtain the solution $x(t)$. The ratio of \eqref{brzinaDHO} and \eqref{polozajDHO} can be written as
\begin{equation}
\frac{1}{x(t)}\frac{dx(t)}{dt}=\omega_0\frac{1}{\tan\phi(t)}\,.
\label{ratio0}
\end{equation}
We insert \eqref{PhiDHO} in \eqref{ratio0} and get
\begin{equation}
\frac{1}{x(t)}\frac{dx(t)}{dt}=-\frac{\omega_0^2}{\omega}\cdot\frac{\sin(\omega t)}{\cos(\omega t)+\frac{\gamma}{\omega}\sin(\omega t)}\,.
\label{ratio1}
\end{equation}
The denominator on the right hand side of \eqref{ratio1} is 
\begin{equation}
D(t)=\cos(\omega t)+\frac{\gamma}{\omega}\sin(\omega t)\,,
\label{nazivnik}
\end{equation}
and its derivative is
\begin{equation}
\frac{dD(t)}{dt}=-\omega\sin(\omega t)+\gamma\cos(\omega t)\,.
\label{dtnazivnik}
\end{equation}
The numerator on the right side of \eqref{ratio1}, i.e. $\sin(\omega t)$, can be written as a linear combination of the denominator $D(t)$ and its derivative, i.e.
\begin{equation}
\sin(\omega t)=\frac{\gamma\omega}{\omega_0^2}D(t)-\frac{\omega}{\omega_0^2}\frac{dD(t)}{dt}\,.
\label{sinD}
\end{equation}
We insert \eqref{sinD} into \eqref{ratio1} and get
\begin{equation}
\frac{1}{x(t)}\frac{dx(t)}{dt}=-\gamma+\frac{1}{D(t)}\frac{dD(t)}{dt}\,.
\label{ratio2}
\end{equation}
Both sides of \eqref{ratio2} can be written as total differentials, i.e.
\begin{equation}
d\left(\ln|x(t)|\right)=d\left(-\gamma t+\ln|D(t)|+const\right)\,,
\label{totaldiff1}
\end{equation}
where we added a constant since $d(const)=0$ holds. We can use $-\gamma t=\ln(e^{-\gamma t})$ in \eqref{totaldiff1}, and write the constant term as $\ln C$, where $C>0$ is a constant. Thus, \eqref{totaldiff1} can be rewritten as 
\begin{equation}
d\left(\ln|x(t)|\right)=d\left(\ln\left|C e^{-\gamma t}D(t)\right|\right)\,,
\label{totaldiff2}
\end{equation}
which gives us
\begin{equation}
|x(t)|=\left|Ce^{-\gamma t}\left(\cos(\omega t)+\frac{\gamma}{\omega}\sin(\omega t)\right)\right|\,,
\label{absX}
\end{equation}
Using the initial condition $x(0)=x_0>0$ we easily get $C=x_0$. Thus, the solution is
\begin{equation}
x(t)=x_0e^{-\gamma t}\left(\cos(\omega t)+\frac{\gamma}{\omega}\sin(\omega t)\right)\,,
\label{XDHOsolution}
\end{equation}
which corresponds to the standard form of the solution \cite{Berkeley}. The velocity corresponding to \eqref{XDHOsolution} is
\begin{equation}
v(t)=\frac{dx(t)}{dt}=-\frac{\omega_0^2x_0}{\omega}e^{-\gamma t}\sin(\omega t)\,.
\label{VDHOsolution}
\end{equation}
Using \eqref{XDHOsolution} and \eqref{VDHOsolution} in $E(t)=mv^2(t)/2+kx^2(t)/2$ we obtain  
\begin{equation}
E(t)=E_0e^{-2\gamma t}\left(1+\frac{\gamma}{\omega}\sin(2\omega t)+2\left(\frac{\gamma}{\omega}\right)^2\sin^2(\omega t)\right)\,,
\label{EDHO}
\end{equation}
where $E_0=m\omega_0^2x_0^2/2$. 

We can easily obtain the solution and energy of the critical damping regime by taking the $\gamma\rightarrow\omega_0$ limit of \eqref{XDHOsolution} and \eqref{EDHO}, and to obtain the solution and energy of the overdamped regime, we can write $\omega=i\sqrt{\gamma^2-\omega_0^2}$ in \eqref{XDHOsolution} and \eqref{EDHO} and use the connection of trigonometric and hyperbolic functions \cite{Berkeley}.

\begin{figure}[h!t!]
\begin{center}
\includegraphics[width=0.45\textwidth]{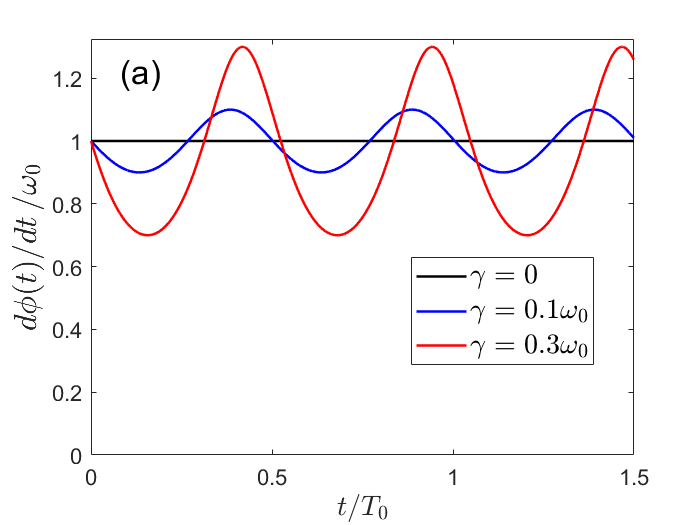}
\includegraphics[width=0.45\textwidth]{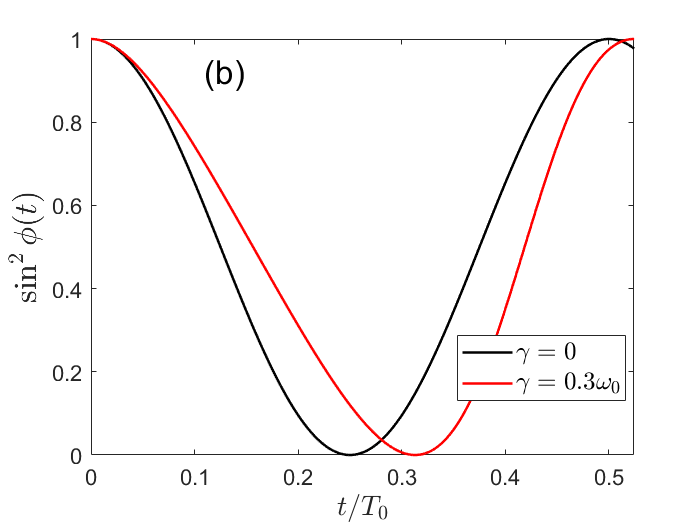}
\end{center}
\caption{(a) For damped harmonic oscillator (with $\gamma>0$), angular velocity $d\phi(t)/dt$ of the phasor $\tilde{R}(t)$ oscillates in time around the corresponding undamped value, i.e. around $\omega_0$. (b) The fraction of potential energy in the total energy, $U(x(t))/E(t)=\sin^2\phi(t)$ for undamped oscillator and damped oscillator with $\gamma=0.3\omega_0$. On both figures, the time unit is $T_0=2\pi/\omega_0$, i.e. the period of the undamped oscillator. See text for details.} 
\label{slika0}
\end{figure}

The corresponding phasor $\tilde{R}(t)=\sqrt{m/2}\,v(t)+i\sqrt{k/2}\,x(t)$ has time-dependent (monotonically decreasing) magnitude $|\tilde{R}(t)|=\sqrt{E(t)}$ and rotates with the angular velocity given by time derivative of \eqref{PhiDHO}, i.e. by
%
%
%
\begin{equation}
\frac{d\phi(t)}{dt}=\frac{\omega_0}{1+\frac{2\gamma}{\omega}\sin(\omega t)\left(\cos(\omega t)+\frac{\gamma}{\omega}\sin(\omega t)\right)}\,.
\label{kutnaDHO}
\end{equation}
Since $d\phi(t)/dt>0$ for all $t\geq0$, we can conclude that $\tilde{R}(t)$ rotates counter-clockwise (starting from the initial phase $\phi(0)=\pi/2$, for our choice of initial conditions). In Fig.\,\ref{slika0}\,(a) we show the angular velocity \eqref{kutnaDHO} for $\gamma/\omega_0=\lbrace0, 0.1, 0.3\rbrace$. We can see that, for $\gamma>0$, the angular velocity can be less than, equal, or greater than $\omega_0$. It is easy to show that \eqref{kutnaDHO} is equal to $\omega_0$ at $t=0$, instants when the system passes through the equilibrium position, and instants when it passes through the turning points. For example, we can confirm this by noting that the trigonometric part of the denominator of the expression \eqref{kutnaDHO} can be related to the product of \eqref{XDHOsolution} and \eqref{VDHOsolution}, i.e. by noting that
\begin{equation}
\sin(\omega t)\left(\cos(\omega t)+\frac{\gamma}{\omega}\sin(\omega t)\right)\propto \frac{v(t)x(t)}{e^{-2\gamma t}}\,
\label{zajebancija}
\end{equation}
holds. Thus, as the damped oscillator moves from the turning point towards the equilibrium position, its angular velocity is less than $\omega_0$, and as it moves from the equilibrium position towards the next turning point, its angular velocity is greater than $\omega_0$. In Fig.\,\ref{slika0}\,(b) we show the potential energy to the total energy ratio as a function of time, i.e. $U(x(t))/E(t)=\sin^2\phi(t)$, for $\gamma/\omega_0=\lbrace 0, 0.3\rbrace$. We see that the damped system (in the underdamped regime) reaches the equilibrium position (instant with $\sin^2\phi(t)=0$) more slowly than the undamped one, but, interestingly, the damped system takes less time to move from the equilibrium position to the turning point (instant with $\sin^2\phi(t)=1$) than the undamped system. This behavior is easy to understand conceptually. For our choice of initial conditions, the dynamics starts with purely potential energy and, due to damping, the damped oscillator reaches the equilibrium position more slowly and with a lesser kinetic energy (lesser speed) than the corresponding undamped oscillator. From the equilibrium position onward, a damped oscillator is slowed down by a spring force and by the damping force; thereby, the transition from the equilibrium position to a turning point takes lesser time with damping than without damping. Of course, in the critically damped and overdamped oscillators, this effect does not exist, since in these cases the system cannot overshoot the equilibrium position (for initial conditions considered here).

\subsection{Approximation of the energy decay and approximate analytical solution in the weak damping regime}
\label{DHO2}

In addition to the simplicity and ease in treating the undamped harmonic oscillator in Section \ref{SHO}, our approach also simplifies the treatment of a weakly damped harmonic oscillator. By the time they come to the topic of the damped harmonic oscillator, undergraduate physics and engineering students have enough knowledge of mathematics and physics to easily understand the derivation of equations \eqref{brzinaDHO}-\eqref{dEDHO}. Students can easily see, for example, by considering demonstration experiments in laboratory exercises or lectures, that systems described by the damped harmonic oscillator model, for weak enough damping, behave like a simple (undamped) harmonic oscillator whose amplitude slowly decreases with time, while the frequency remains practically unchanged with respect to frequency of the undamped oscillator. 

Therefore, even to students without any prior formal knowledge of the damped harmonic oscillator, it is easy to explain that we can take $\phi(t)\approx\omega_0t+\phi_0$ in equations \eqref{brzinaDHO}-\eqref{dEDHO} in the case of weak damping, and derive the approximate analytical expressions for $E(t)$ and $x(t)$. Thus, for weak damping and initial conditions $x_0>0$ and $v_0=0$ (which correspond to $\phi_0=\pi/2$), the approximation of equation \eqref{dEDHO} is
\begin{equation}
\frac{dE(t)}{dt}\approx-4\gamma E(t)\sin^2(\omega_0t)\,.
\label{dEDHOweak}
\end{equation}
We can rewrite \eqref{dEDHOweak} and integrate both sides, i.e. 
\begin{equation}
\int_{E_0}^{E(t)}\frac{dE(t')}{E(t')}\approx-4\gamma\int_0^t\sin^2(\omega_0t')dt'\,.
\label{dEDHOweak1}
\end{equation}
Both integrals in \eqref{dEDHOweak1} are elementary. The solution of the left integral is $\ln(E(t)/E_0)$, and the solution of the right integral is $t/2-\sin(2\omega_0t)/(4\omega_0)$, thus the approximate energy is
\begin{equation}
E_{approx}(t)=E_0\,e^{-2\gamma t}e^{\frac{\gamma}{\omega_0}\sin(2\omega_0t)}\,.
\label{Eapprox}
\end{equation}
Since for weak damping $\gamma/\omega_0\ll1$ holds, we can use $e^{\frac{\gamma}{\omega_0}\sin(2\omega_0t)}\approx1+\frac{\gamma}{\omega_0}\sin(2\omega_0t)$, and \eqref{Eapprox} becomes
\begin{equation}
E_{approx}(t)=E_0\,e^{-2\gamma t}\left(1+\frac{\gamma}{\omega_0}\sin(2\omega_0t)\right)\,.
\label{Eapprox1}
\end{equation}
Thus, the approximation of \eqref{polozajDHO} is
\begin{equation}
x_{approx}(t)=x_0\,e^{-\gamma t}\sqrt{1+\frac{\gamma}{\omega_0}\sin(2\omega_0t)}\,\cos(\omega_0t)\,.
\label{Xapprox0}
\end{equation}
We can further simplify \eqref{Xapprox0}, since $\sqrt{1+\frac{\gamma}{\omega_0}\sin(2\omega_0t)}\approx1+\frac{\gamma}{2\omega_0}\sin(2\omega_0t)$ is valid for $\gamma/\omega_0\ll1$, and get
\begin{equation}
x_{approx}(t)=x_0\,e^{-\gamma t}\left(1+\frac{\gamma}{2\omega_0}\sin(2\omega_0t)\right)\,\cos(\omega_0t)\,
\label{Xapprox}
\end{equation}
as a final expression for our approximate analytical solution. 

\begin{figure}[h!t!]
\begin{center}
\includegraphics[width=0.45\textwidth]{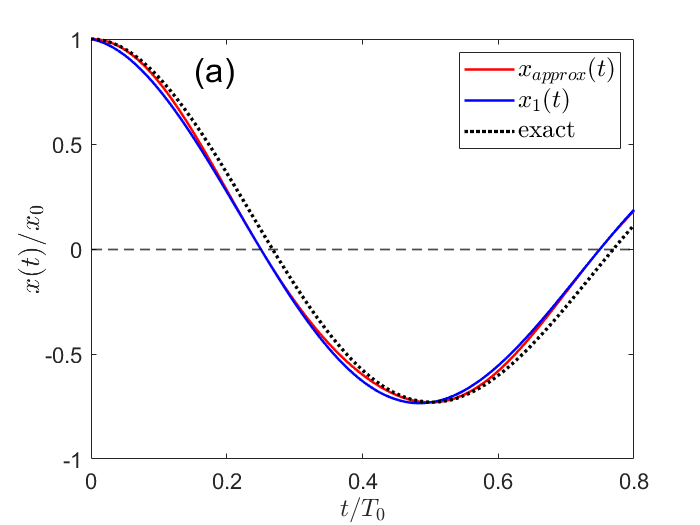}
\includegraphics[width=0.45\textwidth]{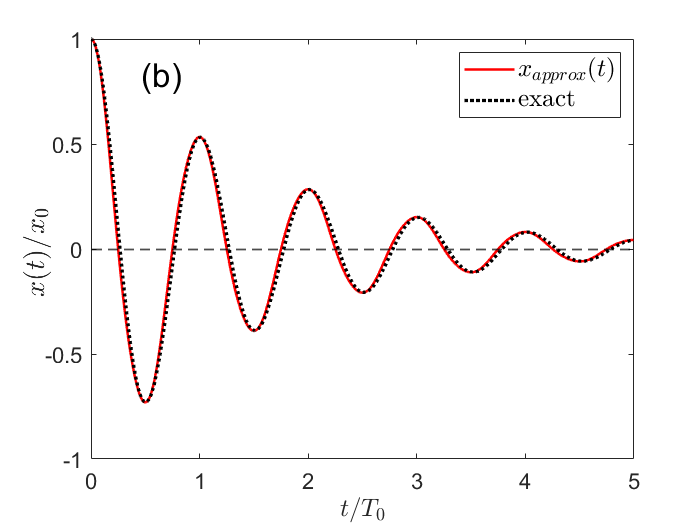}
\end{center}
\caption{Solid red curve corresponds to the approximate solution $x_{approx}(t)$, i.e. \eqref{Xapprox}, solid blue curve to $x_1(t)=x_0e^{-\gamma t}\cos(\omega_0t)$, and black dotted curve to exact analytical solution \eqref{XDHOsolution}. All curves are obtained with $\gamma=0.1\omega_0$, and $T_0=2\pi/\omega_0$ is the period of undamped oscillator. (a) We show a shorter time span in order to see more clearly the better match between $x_{approx}(t)$ and the exact solution in vicinity of the turning points, compared to $x_1(t)$. (b) Comparison of $x_{approx}(t)$ and the exact solution over a larger time span.} 
\label{slika1}
\end{figure}

The approximate energy \eqref{Eapprox1} was recently derived by Lelas et al. \cite{LelasPezer}, in a simple but significantly more involved way, compared to the approach presented here. It has been shown that the approximation \eqref{Eapprox1} excellently overlaps with the exact energy in the weak damping regime  \cite{LelasPezer}, i.e. for $0<\gamma/\omega_0\lesssim0.1$. Therefore, we omit here the analysis of the validity of the approximation \eqref{Eapprox1}, but we will briefly analyze the validity of the approximate solution \eqref{Xapprox}, since it is different from the approximate solution $x_1(t)=x_0e^{-\gamma t}\cos(\omega_0t)$ (we have put the subscript $1$ just for notation) obtained in the paper \cite{LelasPezer}. It is clear that the solution \eqref{Xapprox} and $x_1(t)$ have zero crossings at the same instants, but in Fig.\,\ref{slika1}\,(a) we see that \eqref{Xapprox} overlaps better with the exact solution at the turning points. Therefore, with an approach that is simpler than the one used in the paper \cite{LelasPezer}, we obtained an equally good approximation of the energy and somewhat better approximate analytical solutions.
%


\section{Limitations of our approach}
\label{limitations}

The examples which we treated in the previous sections (the harmonic oscillator $U(x)=kx^{2}/2$, the homogeneous gravitational
potential $U(x)=mgx$, and the inverse-square form $U(x)=K/x^{2}$), share a common feature: after inserting the corresponding $u(x)$ into Eq.\eqref{polozaj} and combining with Eq. \eqref{brzina}, the resulting equation for the phase ${\phi}(t)$ reduces to an elementary integral. In general, this will not be the case. For many 1D conservative systems the phase integral becomes non-elementary, and the solution for $\phi(t)$ (and $x(t)$) cannot be expressed in closed form.

Let us illustrate this by considering the power-law potentials
\begin{equation}
U(x)=\frac{K_{n}}{n}\, x^{n}, \qquad K_{n}>0,
\end{equation}
with $n$ not restricted to integers. Using $U(x)=E\sin^{2}\phi$ (i.e., Eq.\eqref{polozaj}), the position can be formally expressed as
\begin{equation}
x(\phi)=\left(\frac{nE}{K_{n}}\right)^{1/n} \sin^{2/n}\!\phi,
\end{equation}
which, after calculating the time derivative and comparing that with Eq. \eqref{brzina}, leads to the phase equation:
\begin{equation}
\frac{d \phi}{dt}
   = C_{n}\, \sin^{\,1-2/n}\!\phi,
\end{equation}
where $C_{n}= \frac{n}{2}
  \sqrt{\frac{2E}{m}}
  \left(\frac{nE}{K_n}\right)^{-1/n}$ is a constant. The corresponding integral,
\begin{equation}
t
   = \frac{1}{C_n} \int_{\phi_0}^{\phi(t)} \sin^{\,2/n-1}\!\phi d\phi ,
\end{equation}
is elementary only for special exponents. The solvable cases discussed earlier in the paper correspond precisely to the values $n=1,\ n=2,\ n=-2,$ i.e. the linear potential, the harmonic oscillator, and the inverse-square potential.

However, for generic $n$, the integral cannot be expressed using elementary functions; instead, it reduces to an incomplete elliptic integral or a hypergeometric function, and no closed-form expression for $x(t)$ can be found. We emphasize that this is not a limitation of our method: the standard approach based on direct integration of the equation of motion will encounter analogous difficulties. 
In summary, our approach provides a unified and often simpler route to solvable systems, and it also makes clear why many potentials do not admit elementary solutions.


As for damped harmonic oscillators, the other two most common types of damping that appear in physics and engineering are sliding friction \cite{Lapidus, AviAJP, Grk2, Kamela}  (often called Coulomb damping) and damping quadratic in velocity \cite{Smith, Mungan, Wang, Grk2}. These two types of damping can be modeled by a force of the form $F_d(v)=-\text{sgn}(v)\,b_nv^n$ (see, e.g. Ref. \cite{Wang}), where $n=\lbrace0,2\rbrace$, $b_n>0$ are the corresponding damping constants, and $\text{sgn}(v)$ is the sign function (equal to $1$ for $v>0$, $-1$ for $v<0$, and $0$ for $v=0$). Of course, in these two cases, equations \eqref{brzinaDHO} and \eqref{polozajDHO} keep the same form, but equation \eqref{dEDHO}, i.e. the energy dissipation rate $dE/dt=v\cdot F_d(v)$, becomes
\begin{equation}
\frac{dE(t)}{dt}=-b_n\left(\frac{2E(t)}{m}\right)^{\frac{n+1}{2}}|\cos\phi(t)|\cos^n\phi(t)\,.
\label{nonlin_dE}
\end{equation}
We used $\text{sgn}\left(v(t)\right)\cdot v(t)=|v(t)|$ in deriving \eqref{nonlin_dE}. If we continue with the steps that led to the phase integral \eqref{fazaDHOintegral} in the case of linear damping, we get an integral of the form 
\begin{equation}
\int_{\phi_0}^{\phi(t)}\frac{d\phi(t')}{1+\frac{b_n}{m\omega_0}\left(2E(t)/m\right)^{\frac{n-1}{2}}F(\phi(t))}=\omega_0\,t\,,
\label{fazaDHOnonlin}
\end{equation}
where $F(\phi(t))=\sin\phi(t)|\cos\phi(t)|(\cos\phi(t))^{n-1}$. Thus, the energy and phase remained coupled, i.e. our approach cannot provide exact solutions for these two types of damping.
Nevertheless, the approach and steps we used to determine the approximate analytical solution in the case of linear damping, i.e. \eqref{Xapprox}, have the potential to provide new approximate analytical solutions for these two types of damping, as well as for other dissipative systems such as, e.g., vertical projectile motion with air drag. A detailed analysis of the application of our approach to other dissipative systems will be the topic of future work.

\section{Comment on systems with negative definite potential energy}
\label{negativeU}

Our energy factorization approach cannot be directly applied to systems with negative-definite potential energy, i.e. with $U(x)\leq0$, but it is easy to adapt it for such cases as well. We can write the negative-definite potential energy as $U(x)=-u^2(x)$, where again we use real-valued auxiliary function $u(x)$. Thus, the law of conservation of total mechanical energy can be rewritten as 
\begin{equation}
\frac{m}{2E}v^2(t)-\frac{1}{E}u^2(x(t))=1\,.
\label{negativeU1}
\end{equation}
Since the identity $\cosh^2\phi(t)-\sinh^2\phi(t)=1$ holds for any function $\phi(t)$, it is easy to argue that equations
\begin{equation}
v(t)=\sqrt{\frac{2E}{m}}\cosh\phi(t)\,\,\,\text{and}\,\,\,u(x(t))=\sqrt{E}\sinh\phi(t)
\label{Vch}
\end{equation}
%
%
%
hold for $E>0$, while
\begin{equation}
v(t)=\sqrt{\frac{2|E|}{m}}\sinh\phi(t)\,\,\,\text{and}\,\,\,u(x(t))=\sqrt{|E|}\cosh\phi(t)
\label{Vch2}
\end{equation}
%
%
%
hold for $E<0$. These equations are similar to equations \eqref{brzina} and \eqref{polozaj}, except that instead of trigonometric functions, hyperbolic functions now appear. The case $U(x)\leq0$ with $E=0$ is not covered by these equations, but can be treated straightforwardly by standard methods \cite{Goldstein2002}.



\section{Conclusion and outlook}
\label{zakljucak}

We have introduced a new approach to solving several fundamental classical conservative systems and a damped harmonic oscillator, which completely bypasses solving Newton's equations of motion. On the mathematical side, our approach is based on the basic properties of complex numbers and elementary knowledge of calculus, and as such can be extremely useful in undergraduate physics teaching. For example, the (undamped) harmonic oscillator is one of the most fundamental models in physics, and its solution is typically derived, at the undergraduate level, via a connection with uniform circular motion \cite{Cutnell8, UniPhys, Resnick10}. We have shown how to derive the solution of a harmonic oscillator in a couple of easy-to-understand steps starting from the total mechanical energy. Using our approach, it is not necessary to introduce a connection with uniform circular motion, and once they have the solution, students can easily conclude for themselves that this connection exists.

In the case of a harmonic oscillator with linear damping, our derivation of the exact analytical solution is of course somewhat more involved compared to the undamped case, but even in that case we use mathematics that is typically used in undergraduate physics courses anyway. Our approach to the approximate solution of this system is easy to understand and, in addition to an excellent approximation of the energy, gives an approximate analytical solution that, to the best of our knowledge, cannot be found elsewhere in the literature.

We also commented on the limitations of our approach and pointed out for which conservative systems it is applicable and for which it is not, at least not in the sense of obtaining exact analytical solutions. Furthermore, we have shown that our approach can be adapted for negative-definite potential energies as well. In future research, we will apply our approach to the derivation of new approximate analytical solutions for several other dissipative systems, as we have already commented, but also investigate the potential of our approach for finding approximate solutions of conservative systems with potential energies such that $x(t)$ (or $r(t)$ in the case of 3D central forces) cannot be obtained in closed form.

\section{Acknowledgments}

We acknowledge support from the project “Implementation of cutting-edge research and its application as part of the Scientific Center of Excellence for Quantum and Complex Systems, and Representations of Lie Algebras”, Grant No. PK.1.1.10.0004, co-financed by the European Union through the European Regional Development Fund - Competitiveness and Cohesion Programme 2021-2027. D. J. also acknowledges support from the institutional project “TopoPha – Topological Phases” (100‑021/26) of the University of Zagreb Faculty of Civil Engineering, funded by the European Union – NextGenerationEU.



\begin{appendix}
    \section{Solution to the integral \eqref{fazaDHOintegral} and phase $\phi(t)$ for underdamped case}
    \label{appendixA}

The phase integral is
\begin{equation}
\int_{\phi_0}^{\phi(t)}
\frac{d\phi'}{1+(\gamma/\omega_0)\sin(2\phi')}
=
\omega_0 t,
\label{eq:A_phase_integral}
\end{equation}
We use the substitution
\begin{equation}
u = \tan\phi, 
\qquad
d\phi = \frac{du}{1+u^2},
\qquad
\sin(2\phi) = \frac{2u}{1+u^2}.
\end{equation}
Equation~\eqref{eq:A_phase_integral} then becomes
\begin{equation}
\int_{u_0}^{u(t)}
\frac{du}{Q(u)}
=
\omega_0 t,
\label{eq:A_u_integral}
\end{equation}
where $u_0 = \tan\phi_0$, and the denominator is the quadratic
\begin{equation}
Q(u) = u^2 + 2\frac{\gamma}{\omega_0}u + 1.
\label{eq:A_quadratic}
\end{equation}
For $0<\gamma/\omega_0<1$, we can define $\omega=\sqrt{\omega_0^2-\gamma^2}>0$ and
\begin{equation}
\Delta =(\omega/\omega_0)^2=1-(\gamma/\omega_0)^2 > 0\,.
\label{Delta}
\end{equation}
Using \eqref{Delta} we complete the square, i.e.
\begin{equation}
Q(u) = \left(u + \frac{\gamma}{\omega_0}\right)^2 + \Delta\,.
\end{equation}
Thus
\begin{equation}
\int \frac{du}{Q(u)}
=
\frac{1}{\sqrt{\Delta}}
\arctan\!\left(
\frac{u + \gamma/\omega_0}{\sqrt{\Delta}}
\right)
+ C.
\end{equation}
Applying the limits $u_0 \to u(t)$ in Eq.~\eqref{eq:A_u_integral} gives
\begin{equation}
\frac{1}{\sqrt{\Delta}}
\left[
\arctan\!\left(
\frac{u(t) + \gamma/\omega_0}{\sqrt{\Delta}}
\right)
-
\arctan\!\left(
\frac{u_0 + \gamma/\omega_0}{\sqrt{\Delta}}
\right)
\right]
=
\omega_0 t.
\label{eq:A_under_arctan}
\end{equation}
For the initial conditions $x_0>0$ and $v_0=0$, we have
\begin{equation}
\phi_0 = \frac{\pi}{2},
\qquad
u_0 = \tan\phi_0 \to +\infty.
\end{equation}
In this limit,
\begin{equation}
\arctan\!\left(
\frac{u_0 + \gamma/\omega_0}{\sqrt{\Delta}}
\right)
\to \frac{\pi}{2},
\end{equation}
so Eq.~\eqref{eq:A_under_arctan} becomes
\begin{equation}
\arctan\!\left(
\frac{u(t) + \gamma/\omega_0}{\omega/\omega_0}
\right)
=
\frac{\pi}{2} + \omega t\,,
\end{equation}
where we used $\sqrt{\Delta}=\omega/\omega_0$. Taking the tangent of both sides,
\begin{equation}
\frac{u(t) + \gamma/\omega_0}{\omega/\omega_0}
=
\tan\!\left(\frac{\pi}{2} + \omega t\right)
=
-\cot(\omega t),
\end{equation}
so
\begin{equation}
u(t) = \tan\phi(t)
=
-\frac{\omega}{\omega_0}\,\cot(\omega t)
-
\frac{\gamma}{\omega_0}.
\label{eq:A_tanphi_underdamped}
\end{equation}
Thus, the phase is
\begin{equation}
\phi(t)
=
\arctan\!\left[
-\frac{\omega}{\omega_0}\,\cot(\omega t)
-
\frac{\gamma}{\omega_0}
\right]\,.
\label{fazaAppendix}
\end{equation}
%


\end{appendix}
\bibliography{mainEJP2026R1}

\end{document}